\documentclass[reprint,amsmath,amssymb,pra,twocolumn,longbibliography]{revtex4-2}
 \usepackage{amsmath}
 \usepackage{csquotes}
  \usepackage{graphicx}
\usepackage{color}

\def\josab{J. Opt. Soc. Am. B}

\def\ol{Optics Letters}
\def\prb{Phys. Rev. B}

\def\pra{Phys. Rev. A}
\def\prl{Phys. Rev. Lett.}

\newcommand{\eps}{\varepsilon}

\begin{document}
\title{Conditional robustness of propagating bound states in
  the continuum \\ on biperiodic structures}


\author{Lijun Yuan}
\email{Corresponding author: ljyuan@ctbu.edu.cn}
\affiliation{College of Mathematics and Statistics, Chongqing Technology and Business University, Chongqing, 
China \\ Chongqing Key Laboratory of Social Economic and Applied Statistics, Chongqing Technology and Business University, Chongqing, China}

\author{Ya Yan Lu}
\affiliation{Department of Mathematics, City University of Hong Kong, Hong Kong}

\begin{abstract}

For a periodic structure sandwiched between two homogeneous media, a
bound state in the continuum (BIC) is a guided Bloch mode with a
frequency in the radiation continuum.  Optical BICs have found many
applications, mainly because they give rise to resonances with
ultra-high quality factors.  If the periodic structure has a relevant
symmetry, a BIC may have a symmetry mismatch with incoming and
outgoing propagating waves of the same frequency and compatible
wavevectors, and is considered as protected by symmetry. Propagating
BICs with nonzero Bloch wavevectors have been found on many highly
symmetric periodic structures. They are not protected by symmetry in
the usual sense (i.e., there is no symmetry mismatch), but some of
them seem to depend on symmetry for their existence and robustness.
In this paper, we show that the low-frequency propagating BICs (with
only one radiation channel) on biperiodic structures with an inversion
symmetry in the plane of periodicity and a reflection symmetry in the
perpendicular direction are robust to symmetry-preserving structural
perturbations. In other words, a propagating BIC continues its
existence with a slightly different frequency and a slightly different
Bloch wavevector, when the biperiodic structure is perturbed slightly
preserving the inversion and reflection symmetries.  Our study
enhances theoretical understanding for BICs on periodic structures and
provides useful guidelines for their applications. 

\end{abstract}
\maketitle

 
\section{Introduction}
Periodic structures sandwiched between two homogeneous media are
widely used to study optical bound states in the continuum (BICs)
\cite{hsu16,kosh19}.  A BIC on a  periodic structure is 
a special guided mode above the light line. It decays exponentially in
the surrounding homogeneous media, even though 
its frequency and wavevector are compatible with plane waves
propagating to or from  infinity. For periodic structures with a relevant symmetry, 
symmetry-protected BICs may exist due to a symmetry mismatch
\cite{bonnet94,shipman03,padd00,ochiai01,tikh02,shipman07,lee12,hu15}.
BICs unprotected by symmetry in the usual sense (i.e., 
there is no symmetry mismatch) can also exist on periodic
structures 
\cite{port05,mari08,hsu13_2,yang14,zhen14,bulg14b,gan16,li16,ni16,yuan17,yuan17ol,bulg17pra4,jin19,hu20pra,yuan20b}.
Since a BIC can be regarded as a resonant mode with an infinite quality
factor ($Q$ factor), a small perturbation in the wavevector or  the
structure gives rise to resonant modes with arbitrarily large $Q$
factors \cite{yuan18,kosh18,hu18,yuan20}. Optical BICs have found 
applications in  lasing \cite{kodi17}, sensing \cite{romano19,yesi19}, filtering 
\cite{foley14,cui16}, switching \cite{han19}, nonlinear optics
\cite{yuan16,yuan17_2,yuan_siam}, etc. 


Theoretical questions about the existence and robustness of BICs are
important. The existence of symmetry-protected BICs can be
established rigorously \cite{bonnet94,shipman07}. These
BICs are also robust against small structural
perturbations that preserve the relevant symmetry. If the perturbed
structure breaks the symmetry, a symmetry-protected BIC
usually, but not always, turns to a resonant mode
\cite{yuan20,yuan20b}.  The case for BICs without the usual 
symmetry protection is more complicated. Many of these BICs are found
on highly symmetric structures for relatively low frequencies such
that there is only one radiation channel
\cite{port05,mari08,hsu13_2,bulg14b,gan16,yuan17}. It is certainly possible
for BICs to exist on structures
without any symmetry and/or 
for higher frequencies with more than one radiation channels, but they
are more difficult to find, and usually require the tuning of
parameters \cite{bulg14b,yuan20b}. While the numerical results and
experimental evidences for  
BICs unprotected by symmetry are very convincing, to the best of our
knowledge, there is no rigorous proof for their existence. 
Numerical studies also reveal 
that those BICs on highly symmetric periodic structures are robust to changes in parameters, such as the radius of 
air holes or dielectric rods, dielectric constants of the
components, and the thickness of the structure
\cite{port05,zhen14}. It has been realized  that although 
these BICs are not protected by symmetry in the usual sense, symmetry
still plays a key role for their continual existence
\cite{hsu13_2,zhen14}.
The robustness of a BIC (protected or unprotected by symmetry) can
also be studied by its topological properties. 
The topological charge of a BIC may be defined as the winding number of a projected
polarization vector (the major axis for the polarization  ellipse in
general) of the surrounding resonant 
modes \cite{zhen14,bulg17pra4}. 
This definition assumes the absence of
circularly polarized resonant modes near the BIC, which appears to be
related to the symmetry of the structure \cite{fudan,notomi,yin20}.

In an earlier work \cite{yuan17ol}, we analyzed the robustness of BICs
on 2D structures with 
reflection symmetries in both the periodic and perpendicular
directions.   It was proved rigorously that low-frequency
BICs  (unprotected by symmetry, with 
only one radiation channel) are robust to any symmetry-preserving
perturbations.  In this paper, we study BICs on lossless biperiodic
structures sandwiched between two identical  homogeneous media.
Assuming the structure has an inversion symmetry in the plane of periodicity and a
reflection symmetry in the perpendicular direction,
we show that a typical low-frequency propagating BIC (with a nonzero
Bloch wavevector and only one radiation channel) is robust
 against any lossless structural perturbations that preserve the inversion and
 reflection symmetries.  In other words, if the amplitude of a
 symmetry-preserving  perturbation is sufficiently small, the
 perturbed structure has a BIC 
 with a slightly different frequency and a slightly different Bloch
 wavevector. Since the BICs are not robust against arbitrary
 perturbations that break the required symmetries, we call this a
 conditional robustness result.  

The rest of this paper is organized as follows. In Sec.~II, we
describe the biperiodic structure and recall the basic 
equations. In Sec.~III, we construct special diffraction
solutions with desirable symmetry properties. In Sec.~IV, we scale 
the BICs and reveal their symmetry properties. These regularized
diffraction solutions and BICs are used in Sec.~V to establish the
conditional robustness result. In Sec.~VI, we present some numerical
examples. The paper is concluded with a brief discussion in Sec.~VII.

\section{Structures and equations}
\label{sec:struct}
We consider a three-dimensional (3D) isotropic and lossless structure that is periodic in
the $x$ and $y$ directions 
with period $L$, has a finite size $2d$ in the $z$ direction, and is
sandwiched between two identical homogeneous media of dielectric constant $\eps_0$. 
The dielectric function $\eps({\bf x})$, for ${\bf   x}=(x,y,z)$, of
the structure and the surrounding media,  is real, satisfies $\eps({\bf
  x}) = \eps_0$ for $|z| > d$ and 
\begin{equation}
  \label{biperiodic}
 \eps({\bf x}) =   \eps(x + m L, y + nL,z)
\end{equation}
for  all integers $m$ and $n$. In addition to the
periodicity, we assume  the structure has a reflection symmetry in the
$z$-direction and an inversion symmetry in the $xy$ plane, i.e., 
\begin{equation}
\label{eq:symm1} \eps({\bf x}) = \eps(x,y,-z) = \eps(-x,-y,z).
\end{equation}

For isotropic, lossless and non-magnetic 3D structures,  time-harmonic
electromagnetic waves satisfy the following Maxwell's equations 
\begin{eqnarray}
\label{eq:MaxE} \nabla \times E &=& i \omega \mu_0  H \\
\label{eq:MaxH} \nabla \times H &=& - i \omega \epsilon_0 \eps  E, \\
\label{eq:MaxDE} \nabla \cdot  (\eps E) &=& 0, \\
\label{eq:MaxDH} \nabla \cdot H &=& 0,
\end{eqnarray}
where $E$ and $H$ are the electric and magnetic fields respectively,
$\omega$ is 
the angular frequency, $\mu_0$ is the permeability of 
vacuum,  and $\epsilon_0$ is the permittivity of vacuum. 
 The time dependence is assumed to be
$e^{- i \omega   t}$ and is already separated. 
Eliminating $H$,
one obtains  
\begin{equation}
\label{eq:E} \nabla \times \nabla \times E - k^2 \eps E = \nabla (\nabla \cdot E) - \nabla^2 E - k^2 \eps E= 0,
\end{equation}
where $k =\omega/c$ is the freespace wavenumber and $c$ is the 
speed of light in vacuum. If the electric field is known, 
the magnetic field can be easily obtained from Eq.~(\ref{eq:MaxE}).

\section{Diffraction solutions}
\label{sec:diffraction}

In this section, we consider diffraction problems with given incident plane
waves. The main purpose is to construct  diffraction 
solutions with some desirable symmetry properties. These
solutions will be  
used in a perturbation process to prove the conditional robustness of
propagating BICs. 
In the homogeneous media below ($z  < - d$) and above ($z > d$) the
structure, we specify plane incident waves 
\begin{equation}
\label{eq:E_in1} {E}^{(in)}_{\pm}(\mathbf{x})
 = \mathbf{f}^{\pm} e^{i
  \mathbf{k}^{\pm} \cdot \mathbf{x}}, \quad \mp z  > d, 
\end{equation}
where $  \mathbf{k}^{\pm} = (\alpha, \beta, \pm \gamma)$ 
are  real wavevectors satisfying $ \|  \mathbf{k}^{\pm} \|^2 = k^2
\eps_0$ and $\gamma > 0$,  and $\mathbf{f}^{\pm}  = (f_x, f_y, \pm f_z)$ are real 
vectors satisfying $|| {\bf f}^\pm ||=1$. In addition, Eq.~(\ref{eq:MaxDE}) in
the homogeneous media gives the orthogonality condition 
$$\mathbf{f}^{\pm} \cdot \mathbf{k}^{\pm} = 0.$$
We assume the frequency and the wavevector satisfy 
\begin{eqnarray}
\label{eq:One_Channel} 
& &\sqrt{ \alpha^2 + \beta^2} < k \sqrt{\eps_0}    < \nonumber  \\ 
 & & \mbox{min} \left\{ \sqrt{\alpha^2 + \left( \frac{2\pi}{L} - |\beta|
   \right)^2}, \sqrt{ {\beta^2 +} \left( \frac{2\pi}{L} - |\alpha| \right)^2} \right\}. 
\end{eqnarray}
This  implies that the $(0,0)$-th order diffraction channel is the
only propagating channel. More precisely, let 
\begin{eqnarray}
 \hat{\alpha}_j &= &\alpha + 2 j \pi  / L, \quad \hat{\beta}_m = \beta +
 2 m \pi /L, \\
   \hat{\gamma}_{jm} &=& \sqrt{k^2 \eps_0 -
   \hat{\alpha}_j^2 - \hat{\beta}_m^2}
\end{eqnarray}
for all integers $j$ and $m$, where $\hat{\alpha}_0 = \alpha$, $\hat{\beta}_0 = \beta$ and
$\hat{\gamma}_{00} = \gamma$, 
then only $\hat{\gamma}_{00}$ is real and all other
$\hat{\gamma}_{jm}$ for $(j, m)  \neq (0,0)$ are pure imaginary. 

Let $\tilde{E}_e(\mathbf{x}) = [  \tilde{E}_{e,x}({\bf  x}), 
\tilde{E}_{e,y} ({\bf  x}),   \tilde{E}_{e,z}  ({\bf  x})  ]$ 
be a solution of the diffraction problem with incident plane waves
given in Eq.~(\ref{eq:E_in1}). 
Since the structure has a reflection symmetry in the $z$-direction,
the vector field  
\[ 
 \hat{E}_e(\mathbf{x}) = \left[ 
     \tilde{E}_{e,x}(x,y,-z),  \  \tilde{E}_{e,y}(x,y,-z), \
     - \tilde{E}_{e,z}(x,y,-z)  \right]
 \]
also satisfies Eqs.~(\ref{eq:E}) and (\ref{eq:MaxDE}). In addition,
the set of two incident plane waves given in Eq.~(\ref{eq:E_in1}) is
unchanged if we map $z$ to $-z$ and multiply $-1$ to their
$z$-components. Thus, $\hat{E}_e(\mathbf{x})$ solves the
same diffraction problem. If this diffraction problem has a unique
solution, then $\tilde{E}_e(\mathbf{x}) = \hat{E}_e(\mathbf{x})$.
If the diffraction problem does not have a unique solution, we can
still assume $\tilde{E}_e(\mathbf{x}) = \hat{E}_e(\mathbf{x})$,
because otherwise we can replace $\tilde{E}_e(\mathbf{x})$  by $[
  \tilde{E}_e(\mathbf{x}) + \hat{E}_e(\mathbf{x}) ]/2$ which
solves the same diffraction problem. The condition 
$\tilde{E}_e(\mathbf{x}) = \hat{E}_e(\mathbf{x})$ implies that 
the $x$ and $y$
components of $\tilde{E}_e$ are even in $z$ and the $z$ component of
$\tilde{E}_e$ is odd in $z$, i.e., 
\begin{eqnarray}
 \tilde{E}_{e,x}({\bf x}) &=&
\tilde{E}_{e,x}(x,y,-z), \quad 
\tilde{E}_{e,y}({\bf x}) = \tilde{E}_{e,y}(x,y,-z), \nonumber  \\
 \label{eq:scat_even}
  \tilde{E}_{e,z}({\bf x}) &=& - \tilde{E}_{e,z}(x,y,-z).
 \end{eqnarray}
Since the media  for $|z| > d$ are homogeneous and the $(0,0)$-th 
order diffraction channel is the only propagating channel,
$\tilde{E}_e(\mathbf{x})$ has the following asymptotic formula 
\begin{equation}
\label{eq:E_Scat_Full} \tilde{E}_e(\mathbf{x})  \sim  \mathbf{f}^{\mp}
e^{i \mathbf{k}^{\mp} \cdot \mathbf{x}}  +   \mathbf{g}^{\pm} e^{i
  \mathbf{k}^{\pm} \cdot \mathbf{x}} , \quad z \to \pm \infty. 
\end{equation}
where $\mathbf{g}^{\pm}$ are the constant vectors for the outgoing plane
waves. Due to the symmetry given in condition (\ref{eq:scat_even}), the $x$- and
$y$-components of ${\bf g}^\pm$ are identical respectively, and their
$z$-components have opposite signs. In addition, $\mathbf{g}^{\pm}$
must satisfy $\mathbf{g}^{\pm} \cdot \mathbf{k}^{\pm}=0$ due to 
Eq.~(\ref{eq:MaxDE}), and $\| \mathbf{g}^{\pm} \| = 1$ due to energy
conservation. 

Notice that $ \overline{\tilde{E}}_e(-\mathbf{x})$,  i.e. the complex
conjugate of $ \tilde{E}_e(-\mathbf{x})$, also satisfies 
Eqs.~(\ref{eq:E}) and (\ref{eq:MaxDE}), and has the asymptotic formula 
$$ \overline{\tilde{E}}_e(-\mathbf{x}) 
\sim  \overline{\mathbf{g}}^{\mp} e^{i \mathbf{k}^{\mp} \cdot \mathbf{x}}  +  \overline{\mathbf{f}}^{\pm} e^{i \mathbf{k}^{\pm} \cdot \mathbf{x}} , \quad z \to \pm \infty. $$
Therefore,  $\overline{\tilde{E}}_e(-\mathbf{x})$ can be regarded as a
solution of the diffraction problem with incident plane waves
$\overline{\mathbf{g}}^{\mp} e^{i \mathbf{k}^{\mp} \cdot \mathbf{x}}
$. If $\mathbf{f}^- + \overline{\mathbf{g}}^{-} \neq 0$, we let
$E_e(\mathbf{x}) = \tilde{E}_e(\mathbf{x}) +
\overline{\tilde{E}}_e(-\mathbf{x}) $, then $E_e$ satisfies the
parity-time ($\mathcal{PT}$) symmetry condition 
\begin{equation}
\label{eq:scat_pt_xyz} E_e(\mathbf{x}) = \overline{E}_e(-\mathbf{x}).
\end{equation}
The asymptotic formula of $E_e(\mathbf{x})$ at infinity can be written as
\begin{equation}
\label{eq:E_s_e}E_e(\mathbf{x}) \sim \left\{ \begin{matrix} \mathbf{c}^-   e^{i \mathbf{k}^{-} \cdot \mathbf{x}} +  \overline{\mathbf{c}}^+  e^{i \mathbf{k}^+ \cdot \mathbf{x}},  & z \to +\infty, \\
\mathbf{c}^+ e^{i \mathbf{k}^{+} \cdot \mathbf{x}} +  \overline{\mathbf{c}}^- e^{i \mathbf{k}^- \cdot \mathbf{x}},  & z \to -\infty, \end{matrix} \right. 
\end{equation}
where 
$\mathbf{c}^{\pm} = {\mathbf{f}^{\pm} + \overline{\mathbf{g}}^{\pm}
}$. 
It is easy to verify that $\| \mathbf{c}^- \| = \| \mathbf{c}^+
\|$, the $x$- and $y$-components of $\mathbf{c}^{\pm}$ are identical
respectively, and the $z$-components have opposite signs.  We can
scale $E_e$  
such that $ \| \mathbf{c}^{\pm} \| = 1.$  
If $\mathbf{f}^- + \overline{\mathbf{g}}^{-} =0 $, then  $E_e = i
\tilde{E}_e$ is a solution of a diffraction problem with incident
plane waves $i \mathbf{f}^{\pm} e^{i \mathbf{k}^{\pm} \cdot 
  \mathbf{x}}$. In that case,  conditions (\ref{eq:scat_pt_xyz}) and
(\ref{eq:E_s_e}) are still valid with $\mathbf{c}^{\pm} = i
\mathbf{f}^{\pm}$.  In summary, we have constructed a diffraction
solution $E_e$ which satisfies  conditions (\ref{eq:scat_even}) and
(\ref{eq:scat_pt_xyz}). 

Similarly, for incident plane waves 
\begin{eqnarray}
{E}^{(in)}_-(\mathbf{x}) &=& \frac{\mathbf{k}^- \times \mathbf{f^-}}{ \| \mathbf{k}^{-} \times \mathbf{f}^{-} \|} e^{i \mathbf{k}^{-} \cdot \mathbf{x}},  \\ 
{E}^{(in)}_+(\mathbf{x}) &=& - \frac{\mathbf{k}^+ \times \mathbf{f^+}}{ \| \mathbf{k}^{+} \times \mathbf{f}^{+} \|} e^{i \mathbf{k}^{+} \cdot \mathbf{x}},
\end{eqnarray}
given in the media for $z > d$ and $z < -d$, respectively, we can
construct a diffraction solution $E^{(2)}_e(\mathbf{x})$   by
following the same procedure above. The solution $E^{(2)}_e(\mathbf{x})$
satisfies the symmetry conditions (\ref{eq:scat_even}) and 
(\ref{eq:scat_pt_xyz}). 
The asymptotic  formula of $E^{(2)}_e(\mathbf{x})$ at infinity is 
\begin{equation}
\label{eq:E_s_e2}E^{(2)}_e(\mathbf{x}) \sim  \left\{ \begin{matrix} {\mathbf{v}}^-   e^{i \mathbf{k}^{-} \cdot \mathbf{x}} +  \overline{\mathbf{v}}^+  e^{i \mathbf{k}^+ \cdot \mathbf{x}},  & z \to +\infty, \\
\mathbf{v}^+ e^{i \mathbf{k}^{+} \cdot \mathbf{x}} +  \overline{\mathbf{v}}^- e^{i \mathbf{k}^- \cdot \mathbf{x}},  & z \to -\infty, \end{matrix} \right. 
\end{equation} 
where $\mathbf{v}^{\pm}$ are defined by the same procedure as
$\mathbf{c}^{\pm}$ and  scaled such that $\| \mathbf{v}^{\pm} \| = 1$.
It is easy to show that $\mathbf{v}^{\pm} \cdot \mathbf{k}^{\pm} =
0$ and  $\mathbf{v}^{\pm} \cdot \mathbf{c}^{\pm} = 0$. Therefore, the 
vectors $\mathbf{k}^+,  \mathbf{c}^+, \mathbf{v}^+ $ form an
orthonormal basis in the 3D space.

If we replace $\mathbf{f}^-$ by $- \mathbf{f}^-$ and follow the same
procedure above,  we can construct diffraction solutions
$E_o(\mathbf{x})$ and $E^{(2)}_o(\mathbf{x})$  that satisfy the same 
$\mathcal{PT}$-symmetry condition (\ref{eq:scat_pt_xyz}) and the 
following condition
\begin{eqnarray}
\label{eq:scat_odd} E_{o,x}({\bf x}) &=& - E_{o,x}(x,y,-z), \quad
E_{o,y}({\bf x}) = - E_{o,y}(x,y,-z), \nonumber \\
\label{eq:scat_odd}  E_{o,z}({\bf x}) &=&  E_{o,z}(x,y,-z).  
 \end{eqnarray}
In other words, the $x$- and $y$-components of $E_o$ (or $E_o^{(2)}$)
are odd in $z$ and the $z$-component of $E_o$ (or $E_o^{(2)}$) is even
in $z$. 

Since the structure is periodic in $x$ and $y$ with
period $L$, a diffraction solution can be written as a Bloch wave
\begin{equation}
\label{eq:def_b}  
E_e(\mathbf{x}) = e^{i \mathbf{b} \cdot \mathbf{x}} \Psi_e(\mathbf{x})
\quad \mbox{for} \quad \mathbf{b} = ( \alpha,   \beta,  0),
\end{equation}
where $\Psi_e(\mathbf{x})$ is periodic in $x$ and $y$ with period $L$
and satisfies the same symmetry conditions as $E_e(\mathbf{x})$,
i.e. (\ref{eq:scat_even}) and (\ref{eq:scat_pt_xyz}). In terms of
$\Psi_e(\mathbf{x})$, the governing equations (\ref{eq:E}) and
(\ref{eq:MaxDE}) become 
\begin{eqnarray}
\label{eq:Phi} \left( \nabla + i \mathbf{b} \right) \times \left( \nabla + i \mathbf{b} \right) \times \Psi_e - k^2 \eps \Psi_e = 0, \\
\label{eq:DPhi} \left( \nabla + i \mathbf{b} \right) \cdot \left(\eps \Psi_e \right) = 0.
\end{eqnarray}
Similarly, we can introduce functions $\Psi_o(\mathbf{x})$,  $\Theta_e(\mathbf{x})$ and $\Theta_o(\mathbf{x})$  such that
\begin{eqnarray}
 E_o(\mathbf{x}) &=& e^{i \mathbf{b} \cdot \mathbf{x}} \Psi_o(\mathbf{x}), \quad E^{(2)}_e(\mathbf{x}) = e^{i \mathbf{b} \cdot \mathbf{x}} \Theta_e(\mathbf{x}), \nonumber \\
 E^{(2)}_o(\mathbf{x}) &=& e^{i \mathbf{b} \cdot \mathbf{x}} \Theta_o(\mathbf{x}).
\end{eqnarray}
These functions are periodic in $x$ and $y$ with period
$L$,  satisfy Eqs.~(\ref{eq:Phi}) and (\ref{eq:DPhi}), and  the same
symmetry conditions as $E^{(2)}_e(\mathbf{x})$, $E_o(\mathbf{x})$ and
$E^{(2)}_o(\mathbf{x})$, respectively.


\section{Bound states in the continuum}
 A Bloch mode on a biperiodic structure sandwiched between two
 homogeneous media is a solution of Eqs.~(\ref{eq:E}) and
 (\ref{eq:MaxDE}) given as 
 \begin{equation}
 \label{eq:Bloch} E(\mathbf{x}) = e^{i (\alpha x +  \beta y)}
 \Phi(x,y,z) = e^{i \mathbf{b} \cdot \mathbf{x}} \Phi(\mathbf{x}), 
 \end{equation}
where $\Phi(\mathbf{x})$ is periodic in $x$ and $y$ with
period $L$ and satisfies Eqs.~(\ref{eq:Phi}) and (\ref{eq:DPhi}), and
${\bf b} = (\alpha, \beta, 0)$ is the Bloch wavevector. A Bloch mode is a
guided (or localized) mode if $(\alpha, \beta)$ is a real pair and
$\Phi \to 0$ as $|z| \to \infty$. Typically, guided modes  that
depend on $(\alpha, \beta)$ and $\omega$ continuously can be found 
below the light cone, i.e. for $ k \sqrt{\eps_0} < \sqrt{\alpha^2 +
  \beta^2}$. Guided modes may also exist in the light cone, i.e. for
$ k \sqrt{\eps_0} > \sqrt{\alpha^2 + \beta^2}$. This is true especially when the
structure has certain symmetry. Such a guided mode in the light cone
is a bound states in the continuum (BIC).    

Since the media for $|z| > d$ are homogeneous, a Bloch mode can be
expanded as 
\begin{equation}
\label{eq:E_plane} E(\mathbf{x}) =
\sum\limits_{j,m=-\infty}^{+\infty} \mathbf{a}^{\pm}_{jm} e^{i
  \mathbf{k}_{jm}^{\pm} \cdot \mathbf{x}},   \quad |z| > d, 
\end{equation}
where the ``$+$'' and ``$-$'' signs correspond to $z > d$ and $z <
-d$, respectively, and 
\begin{equation}
 \mathbf{k}^{\pm}_{jm} = \left( \begin{matrix} \hat{\alpha}_j, \
     \hat{\beta}_m, \ \pm \hat{\gamma}_{jm}  \end{matrix} \right). 
 \end{equation}
 Equation (\ref{eq:MaxDE}) requires that $\mathbf{k}^{\pm}_{jm} \cdot
 \mathbf{a}^{\pm}_{jm} =0$ for all integers $j$ and $m$. If $\omega$
 is real and $\alpha_j^2 + \beta_m^2 > k^2 \eps_0$, then
 $\hat{\gamma}_{j,m}$ is pure imaginary, and the corresponding plane
 wave is  evanescent. For a BIC, all coefficients $\mathbf{a}^{\pm}_{jm} $
 corresponding to real $\hat{\gamma}_{jm}$ must vanish, since the
 field must decay to zero as $|z| \to \infty$. If
 condition~(\ref{eq:One_Channel})
 is satisfied, only $\gamma = \hat{\gamma}_{00}$ is
 real 
 and all other $\hat{\gamma}_{jm}$ for $(j,m) \neq (0,0)$ are pure
 imaginary. In that case, the Bloch mode is a BIC if and only if  $\mathbf{a}^{\pm}_{00} = \mathbf{0}.$

On biperiodic structures with an inversion symmetry in the $xy$
plane, i.e. $\eps({\bf x}) = \eps(-x,-y,z)$, there may exist 
symmetry-protected BICs with $\alpha = \beta = 0$,  and they satisfy  
\begin{eqnarray}
E_x({\bf x}) &=& -E_x(-x,-y,z), \quad E_y({\bf x}) =
-E_y(-x,-y,z),  \nonumber \\
\label{eq:SPBIC_xy}  E_z({\bf x}) &=& E_z(-x,-y,z).  
 \end{eqnarray}
The above condition forces the $x$- and $y$-components of
 $\mathbf{a}^{\pm}_{00}$ to vanish, but since $\mathbf{k}^{\pm}_{00}
 \cdot  \mathbf{a}^{\pm}_{00} =0$, the $z$-components of
 $\mathbf{a}^{\pm}_{00}$ are zero. Therefore, if
 condition~(\ref{eq:One_Channel}) is satisfied, a Bloch mode (with
 $\alpha=\beta=0$) satisfying condition~(\ref{eq:SPBIC_xy}) is always
 a BIC. Notice that the reflection symmetry in $z$ is not required for
 the existence of these symmetry-protected BICs. 

For propagating BICs, we assume condition~(\ref{eq:symm1}) is
satisfied, i.e., the 
structure has an inversion symmetry in the $xy$ plane and a reflection
symmetry in $z$. If $E(\mathbf{x})$ is a propagating BIC,  then  
\begin{equation}
\label{eq:E_tilde} 
\hat{E}(\mathbf{x}) = \left[ \begin{matrix} E_x(x,y,-z),  \
    E_y(x,y,-z), \ - E_z(x,y,-z)  \end{matrix} \right]
\end{equation}
is also a BIC with the same Bloch wavevector and the same frequency.
We can assume the BIC satisfies either 
\begin{eqnarray}
E_x({\bf x}) &=& E_x(x,y,-z), \quad E_y({\bf x}) =
E_y(x,y,-z), \nonumber \\
\label{eq:bic_even_z}  E_z({\bf x}) &=& -E_z(x,y,-z), 
\end{eqnarray} 
or
\begin{eqnarray}
E_x({\bf x}) &=& - E_x(x,y,-z), \quad E_y({\bf x})
= - E_y(x,y,-z), \nonumber \\
\label{eq:bic_odd_z}  E_z({\bf x}) &=& E_z(x,y,-z), 
\end{eqnarray} 
since otherwise, it can be replaced by 
from $[ E(\mathbf{x}) + \hat{E}(\mathbf{x}) ]/2$ or $[  E(\mathbf{x})
- \hat{E}(\mathbf{x}) ]/2$. Notice that the 
vector function $\Phi({\bf x})$ given in Eq.~(\ref{eq:Bloch}) also
satisfies (\ref{eq:bic_even_z}) or (\ref{eq:bic_odd_z}). 

If $E({\bf x})$ is a BIC, it is easy to show that 
$\overline{{E}}(-\mathbf{x})$ is also a BIC with the same frequency
and the same Bloch wavevector. Assuming the BIC 
is non-degenerate (i.e. single), then there must be a constant  $\rho$
such that $E(\mathbf{x}) = \rho \overline{E}(-\mathbf{x})$. 
Evaluating the energy of the BIC on one period of the structure, we
conclude that $\rho$ must satisfy $|\rho| = 1$.  Let $\rho = e^{2 i
  \theta }$, then $W({\bf x}) = e^{- i \theta} E(\mathbf{x})$ is also a BIC and
${W}(\mathbf{x}) = \overline{{W}}(-\mathbf{x})$. Therefore,
without loss of generality, we can assume the propagating BIC
satisfies 
\begin{equation}
\label{eq:E_PT} {E}(\mathbf{x}) = \overline{{E}}(-\mathbf{x}), 
\end{equation}
i.e., it is ${\cal PT}$-symmetric. In that case, the vector function
$\Phi(\mathbf{x})$ given in Eq.~(\ref{eq:Bloch}) is also
$\mathcal{PT}$-symmetric.



\section{Conditional robustness of  propagating BICs}
\label{sec:robust}
In this section, we establish a conditional robustness result for some 
propagating BICs. The robustness of a
BIC refers to its continual existence  under small structural
perturbations. It should be emphasized that the robustness is only
conditional, because there are 
conditions on the original biperiodic structure, the structural perturbation, and
the BIC itself. More specifically,  the biperiodic structure is
required to satisfy the conditions specified in
Sec.~\ref{sec:struct}. Importantly, it must have an inversion symmetry in the 
$xy$ plane and a reflection symmetry along the $z$ axis. The BIC must
be non-degenerate, must have
a Bloch wavevector ${\bf b} =
(\alpha, \beta, 0) \neq {\bf 0}$ and a frequency $\omega$  (or freespace
wavenumber $k$) satisfying condition
(\ref{eq:One_Channel}), and must satisfy $\det({\bf A}) \ne 0$ for the matrix
${\bf A}$ given below. 
Without loss of generality, we 
assume the BIC satisfies symmetry condition (\ref{eq:bic_even_z}), that is, its $x$
and $y$ components are even in $z$ and its $z$ component is odd in 
$z$. 
The dielectric function of a perturbed structure is  given by 
\begin{equation}
\label{eq:Pert_eps} \tilde{\eps}({\bf x}) = \eps({\bf x}) + \delta s(\mathbf{x}), 
\end{equation}
where $ \delta $ is a small real number, $s(\mathbf{x})$ is any $O(1)$
real function satisfying $s(\mathbf{x}) = 0$ for $|z| > d$, the
periodic condition (\ref{biperiodic}) and the symmetry condition
(\ref{eq:symm1}). Under these conditions, we claim that for any
sufficiently small $\delta$, the perturbed structure has a BIC with a
frequency $\tilde{\omega}$ near $\omega$ and a Bloch wavevector
$\tilde{\bf b} = (\tilde{\alpha}, \tilde{\beta}, 0)$ near ${\bf
  b}$. Although the perturbation profile $s({\bf x})$ must preserve
the periodicity and the inversion and reflection symmetries, it can
still be quite arbitrary, therefore, our robustness result is a 
general result. 

To establish the conditional robustness, we construct a BIC on the
perturbed structure using a perturbation method. Let 
$E({\bf x}) = e^{ i (\alpha x+\beta y)} \Phi({\bf x})$  be a BIC on
the original biperiodic structure,  where 
$\Phi({\bf x})$ is periodic in  $x$ and $y$ with period $L$ and tends 
to zero  exponentially as $|z| \to \infty$, 
we look for a BIC $\tilde{E}(\mathbf{x}) = e^{i (\tilde{\alpha} x+
  \tilde{\beta} y)} \tilde{\Phi}(\mathbf{x})$ on the perturbed
structure by expanding 
$\tilde{\Phi}$, $\tilde{k}$, $\tilde{\alpha} $ and $ \tilde{\beta}$ in
power series of $\delta$ 
\begin{eqnarray}
\label{eq:Pert_Phi}  \tilde{\Phi}({\bf x})
&=& \Phi ({\bf x}) + \delta \Phi_1 ({\bf x})+ \delta^2 \Phi_2 ({\bf x}) + \ldots, \\
\label{eq:Pert_k} \tilde{k} &=& k + \delta k_1 + \delta^2 k_2 + \ldots, \\
\label{eq:Pert_alpha} \tilde{\alpha} &=& \alpha + \delta \alpha_1 + \delta^2 \alpha_2 + \ldots, \\
\label{eq:Pert_beta} \tilde{\beta} &=& \beta + \delta \beta_1 +
                                       \delta^2 \beta_2 + \ldots. 
\end{eqnarray} 
The last two expansions above can be written as 
\begin{eqnarray}
\label{eq:Pert_b}  \tilde{\mathbf{b}} &=& \mathbf{b} + \delta 
                                          \mathbf{b}_1 + \delta^2 
                                          \mathbf{b}_2 + \ldots 
\end{eqnarray} 
where ${\bf b}_j = (\alpha_j, \beta_j, 0)$ for $j \ge 1$. 
In the following, we show that for each  $j \ge 1$, $\Phi_j({\bf x})$ can be solved, it
is periodic in $x$ and $y$ with period $L$ and decays to zero 
exponentially as $z \to \infty$,  $k_j$, $\alpha_j$ and $\beta_j$ can
be determined and they are all real numbers.

Substituting expansions (\ref{eq:Pert_Phi}), (\ref{eq:Pert_k}) and
(\ref{eq:Pert_b}) into Eqs.~(\ref{eq:Phi}) and (\ref{eq:DPhi}), and
comparing the coefficients of $\delta^j$ for $j\ge 1$, we obtain the
following equations for $\Phi_j$:
 \begin{eqnarray}
 \label{eq:Phi_j3} & & \mathcal{L} \Phi_j = \alpha_j \mathcal{B}_{1} \Phi + \beta_j \mathcal{B}_{2} \Phi + 2 k k_j \eps \Phi +  F_{j}, \\
\label{eq:Phi_j4} &&  \left(\nabla + i \mathbf{b} \right) \cdot \left(\eps \Phi_j \right) = \alpha_j p_{1} + \beta_j p_{2} + g_j, 
\end{eqnarray}
where 
\begin{eqnarray} 
\label{eq:def_L} \mathcal{L} & =&   \left( \nabla + i \mathbf{b} \right) \times \left( \nabla + i \mathbf{b} \right) \times  - k^2 \eps, \\
\label{eq:Bm} \mathcal{B}_m & = & -i \left[ (\nabla + i \mathbf{b}) \times \mathbf{e}_m \times + \mathbf{e}_m \times (\nabla + i \mathbf{b}) \times \right], \\
p_{m} & =&  -i \eps \mathbf{e}_m \cdot \Phi 
\end{eqnarray}
for $m=1$ and 2, $\mathbf{e}_1=(1,0,0)$ and $\mathbf{e}_2=(0,1,0)$
are unit vectors along the $x$ and $y$ axes, respectively, and 
\[
F_1 = s k^2 \Phi,  \quad g_1 = - \nabla \cdot \left( s \Phi \right), 
\]
and for $j > 1$, 
\begin{eqnarray*}
 &&F_j = - i \sum\limits_{m=1}^{j-1} \left[ \nabla \times \left( \mathbf{b}_{j-m} \times \Phi_m \right) + \mathbf{b}_{j-m} \times \left( \nabla \times \Phi_m \right) \right]  \\ 
& & + \sum\limits_{m=1}^{j-1} \sum\limits_{n=0}^{j-m} \mathbf{b}_{j-m-n} \times \left( \mathbf{b}_n \times \Phi_m \right)  + \sum\limits_{m=1}^{j-1} \mathbf{b}_{j-m} \times \left(\mathbf{b}_m \times \Phi \right) \\
&& + \sum\limits_{m=1}^{j-1} \left[ \eps \sum\limits_{n=0}^{j-m} k_{j-m-n} k_n + s \sum\limits_{n=0}^{j-m-1} k_{j-m-n-1} k_n \right] \Phi_m \\
 & & + \eps \sum\limits_{m=1}^{j-1} k_{j-m} k_m \Phi + s \sum\limits_{m=0}^{j-1} k_{j-m-1} k_m \Phi, 
 \end{eqnarray*}
 \begin{equation*}
 g_j  =  - \nabla \cdot \left( s \Phi_{j-1} \right) - i s \mathbf{b}_{j-1} \cdot  \Phi   - i \sum\limits_{n=1}^{j-1} \left( \eps \mathbf{b}_{j-n}   + s \mathbf{b}_{j-n-1}   \right)\cdot \Phi_n 
 \end{equation*}
In the above,  $\alpha_0 = \alpha$,  $\beta_0 = \beta$, 
$k_0 = k$ and  $\Phi_0 = \Phi$. Notice that 
$\mathcal{L}$, $\mathcal{B}_1$ and $\mathcal{B}_2$ are
operators, $p_1$ and $p_2$ are scalar functions, and all of them are
independent of $j$. Moreover, $F_j$ is a vector function, $g_j$ is a
scalar function, and they do not involve $\alpha_j, \beta_j, k_j$ and
$\Phi_j$. 

%

In the $j$-th step,  we need to determine $\alpha_j$, $\beta_j$ and $k_j$,
and a vector function $\Phi_j$ which  is periodic in $x$ and $y$ and decays to zero
exponentially as $|z| \to \infty$. 
First, we show that if Eqs.~(\ref{eq:Phi_j3}) and (\ref{eq:Phi_j4})
have such a solution $\Phi_j$, then  
$\alpha_j$, $\beta_j$ and $k_j$ must satisfy the following linear system 
\begin{equation}
\label{eq:Linear_sys} { \bf A}  \left[ \begin{matrix}  \alpha_j \\ \beta_j \\ k_j  \end{matrix}  \right] = \left[ \begin{matrix}  a_{11} & a_{12} & a_{13} \\ a_{21} & a_{22} & a_{23} \\ a_{31} & a_{32} & a_{33}  \end{matrix}  \right]  \left[ \begin{matrix}  \alpha_j \\ \beta_j \\ k_j  \end{matrix}  \right] = \left[ \begin{matrix}  b_{1j} \\ b_{2j} \\ b_{3j}  \end{matrix}  \right], 
\end{equation}
where 
\begin{eqnarray}
&& a_{1m} = \int_{\Omega} \overline{\Phi} \cdot \mathcal{B}_{m} \Phi d {\bf x}, \quad a_{13} = 2k  \int_{\Omega} \eps \overline{\Phi} \cdot \Phi d {\bf x}, \\
&&  a_{2m} = \int_{\Omega} \overline{\Psi}_e \cdot \mathcal{B}_{m} \Phi d {\bf x}, \quad a_{23} = 2k  \int_{\Omega} \eps \overline{\Psi}_e \cdot \Phi d {\bf x}, \\
&& a_{3m} = \int_{\Omega} \overline{\Theta}_e \cdot \mathcal{B}_{m} \Phi d {\bf x}, \quad a_{33} = 2k  \int_{\Omega} \eps \overline{\Theta}_e \cdot \Phi d {\bf x},
\end{eqnarray}
for $m=1,2$,  $\Psi_e$ and $\Theta_e$ are related to diffraction
solutions $E_e$ and $E_e^{(2)}$ introduced in Sec.~\ref{sec:diffraction}, 
\begin{eqnarray} 
b_{1j} &=& -\int_{\Omega} \overline{\Phi} \cdot F_j d {\bf x}, \quad b_{2j} = -\int_{\Omega} \overline{\Psi}_e \cdot F_j d {\bf x}, \\
 b_{3j} &=& -\int_{\Omega} \overline{\Theta}_e \cdot F_j d {\bf x},
\end{eqnarray}
and $\Omega$ is the 3D domain given by $|x| < L/2$, $|y| < L/2$ and
$|z| < + \infty$. 
This linear system is obtained
by computing the dot products of Eq.~(\ref{eq:Phi_j3}) with 
$\overline{\Phi}$, $\overline{\Psi}_e$ and $\overline{\Theta}_e$,
  respectively, 
integrating the results on domain $\Omega$, and showing that left hand
sides are all zero (as in  Appendix A). 
Therefore, the three equations in system (\ref{eq:Linear_sys}) are actually
 \begin{eqnarray}
&&  \label{eq:A1} \int_{\Omega} \overline{\Phi} \cdot \left( \alpha_j 
 \mathcal{B}_{m} \Phi + \beta_j \mathcal{B}_{m} \Phi + 2 k k_j \eps 
 \Phi +  F_{j} \right) =0,   \\
&&  \label{eq:A2} \int_{\Omega} \overline{\Psi}_e \cdot \left( \alpha_j 
 \mathcal{B}_{m} \Phi + \beta_j \mathcal{B}_{m} \Phi + 2 k k_j \eps 
 \Phi +  F_{j} \right) =0,   \\
&&  \label{eq:A3} \int_{\Omega} \overline{\Theta}_e \cdot \left(
 \alpha_j \mathcal{B}_{m} \Phi + \beta_j \mathcal{B}_{m} \Phi + 2 k
 k_j \eps \Phi +  F_{j} \right) =0. 
 \end{eqnarray}

Although $\alpha_j$, $\beta_j$ and $k_j$ can be solved from system
(\ref{eq:Linear_sys}) if $\det( {\bf A}) \ne 0$, it is still necessary
to show that Eqs.~(\ref{eq:Phi_j3}) and (\ref{eq:Phi_j4}) indeed have
a solution $\Phi_j$. 
In the following, we show that if ${\bf A}$ is invertible, then
$\alpha_j$, $\beta_j$ and $k_j$ are real, and Eqs.~(\ref{eq:Phi_j3}) and (\ref{eq:Phi_j4})
have a solution $\Phi_j$ that is periodic in $x$
and $y$ with period $L$, decays exponentially as $|z| \to \infty$, is
${\cal PT}$-symmetric, and satisfies the same symmetry condition in
$z$  [assumed to be (\ref{eq:bic_even_z})]  as the BIC. 

For the case $j=1$, since $\Phi$, $\Psi_e$, $\Theta_e$, $\mathcal{B}_m
\Phi$ and $F_1 = s k^2 \Phi$ are all $\mathcal{PT}$-symmetric, the
coefficient matrix ${\bf A}$ and the right hand side of  linear system
Eq.~(\ref{eq:Linear_sys}) are real. Therefore, if $\det({\bf A}) \ne
0$, $\alpha_1, \beta_1$ and $k_1$ can be uniquely solved from
Eq.~(\ref{eq:Linear_sys}),  and they  are real. 
The BIC satisfies ${\cal L} \Phi = 0$, thus the inhomogeneous equation
  (\ref{eq:Phi_j3}) for  $\Phi_j$ is singular. In general, such a
  singular   inhomogeneous equation does not have a solution unless its right
  hand side is orthogonal with the nullspace of ${\cal L}$. 
We have assumed that the BIC is non-degenerate. Therefore,
Eq.~(\ref{eq:Phi_j3}) has solutions if its 
right hand side is orthogonal with $\Phi$, that is, if condition
(\ref{eq:A1}) is satisfied. For the case $j=1$, it is clear that the
right hand side decays to zero as $|z| \to \infty$. Therefore, it is
natural to require $\Phi_1$ to satisfy outgoing radiation
condition as $z \to \pm \infty$.
Since there is only one opening diffraction channel, 
$\Phi_1$ has an asymptotic formula  at infinity
\begin{equation}
\label{eq:Phi_1}  \Phi_1({\bf x}) \sim  \mathbf{d}^{\pm} e^{ \pm i \gamma z},
\quad z \to \pm \infty, 
\end{equation}
where $\mathbf{d}^{\pm}$ are constant vectors. Since the BIC satisfies
the symmetry condition (\ref{eq:bic_even_z}),  the
right hand side of Eq.~(\ref{eq:Phi_j3}) for $j=1$ also satisfies
(\ref{eq:bic_even_z}), and thus  we can assume $\Phi_1$ also
satisfies that condition.  Therefore,  the $x$- and 
$y$-components of $\mathbf{d}^{\pm}$ are identical, respectively, and
their $z$-components have opposite signs. 

To show that $\Phi_1$ decays to zero exponentially as $|z| \to
+\infty$, we only need to show  $\mathbf{d}^{\pm} = {\bf 0}$.  We
proceed by taking
the dot product  of $\overline{\Psi}_e$ with Eq.~(\ref{eq:Phi_j3})
and integrating the result on domain 
\[
\Omega_h = \left\{(x,y,z) \, :   \, 
  |x|<L/2, \, |y|<L/2, \, |z| < h \right\}
\]
 for $h > d$.
Using the asymptotic formulae of $\Phi_1$ and $\Psi_e$ at infinity, 
we can establish the following result
\begin{equation}
\label{eq:coeff_phi1} \lim\limits_{h \to \infty } \int_{\Omega_h} \overline{\Psi}_e \cdot \mathcal{L} \Phi_1 d {\bf x}  = - 4 i \gamma L^2  \mathbf{c}^+ \cdot \mathbf{d}^+. 
\end{equation}
A detailed derivation of Eq.~(\ref{eq:coeff_phi1}) is given in
Appendix B. On the other hand, 
 according to the second equation of system (\ref{eq:Linear_sys}), or Eq.~(\ref{eq:A2}), 
\begin{equation}
\lim\limits_{h \to \infty } \int_{\Omega_h} \overline{\Psi}_e \cdot \mathcal{L} \Phi_1 d {\bf x} = 0.
\end{equation}
  Therefore, we must have
\begin{equation}
\mathbf{c}^+ \cdot \mathbf{d}^+ = 0.
\end{equation}
Similarly, taking the dot product of  $\overline{\Theta}_e$  with
Eq.~(\ref{eq:Phi_j3}), integrating the result in domain $\Omega_h$, and 
letting $h\to \infty$,  we obtain 
\begin{equation}
\mathbf{v}^+ \cdot \mathbf{d}^+ = 0.
\end{equation}
In addition, in the homogeneous medium for $|z|>d$,  Eq.~(\ref{eq:Phi_j4})
  leads to $\mathbf{k}^+ \cdot \mathbf{d}^+ = 0.$ 
From Sec~\ref{sec:diffraction},  we know that 
$\{ \mathbf{k}^+,  \mathbf{c}^+, \mathbf{v}^+ \}$ is an 
  orthonormal basis. Therefore,  we must have $\mathbf{d}^+  =
  \mathbf{d}^-  =   \mathbf{0}$, and thus $\Phi_1$ decays to zero exponentially as
  $|z| \to +\infty.$ 

Meanwhile, since  $\alpha_1$, $\beta_1$ and $k_1$ are real, it is
 easy to verify that the right hand side of Eq.~(\ref{eq:Phi_j3}) for
  $j=1$ is $\mathcal{PT}$-symmetric. 
We can assume $\Phi_1$ is also $\mathcal{PT}$-symmetric, since
otherwise we can replace it  by $\left[ 
    \Phi_1(\mathbf{x}) + \overline{\Phi}_1(-\mathbf{x})\right]/2$
  which is also  a solution of Eqs.~(\ref{eq:Phi_j3}) and
  (\ref{eq:Phi_j4}).  

The same reasoning is applicable to all perturbation steps for $j \geq 
2$.  More specifically,  if ${\bf A}$ is invertible and the
perturbation profile  $s({\bf x})$ satisfies symmetry condition (\ref{eq:symm1}),  and if 
for all $n < j$, $\alpha_n$, $\beta_n$ and $k_n$ are real, and $\Phi_n$ decays
to zero exponentially as $|z| \to \infty$, satisfies symmetry
condition (\ref{eq:bic_even_z}), and is $\mathcal{PT}$-symmetric, 
then we can show that  $\alpha_j$, $\beta_j$ and $k_j$ are real, and
$\Phi_j$ decays to zero exponentially as $|z| \to \infty$, satisfies
condition (\ref{eq:bic_even_z}), and is $\mathcal{PT}$-symmetric. 

If the perturbation profile $s({\bf x})$ does not satisfies condition 
Eq.~(\ref{eq:symm1}), the above perturbation process is likely to fail. In
the first step ($j=1$), in order to have a real $b_{21}$, we need to
have a real $\int_{\Omega} s({\bf x}) \overline{\Psi}_e \cdot \Phi d {\bf x}$. This implies that 
\begin{equation}
\label{eq:s_sym2} \int_{\Omega} \left[ s(\mathbf{x}) - s(-\mathbf{x})
\right] \overline{\Psi}_e \cdot \Phi \, d {\bf x}  = 0.
\end{equation}
Similarly, $s(\mathbf{x})$ should satisfy
\begin{equation}
\label{eq:s_sym3} \int_{\Omega} \left[ s(\mathbf{x}) - s(-\mathbf{x})
\right] \overline{\Theta}_e \cdot \Phi \, d {\bf x}  = 0.
\end{equation}
Moreover, Eqs.~(\ref{eq:A2}) and (\ref{eq:A3}) should still hold  when
$\Psi_e$ and $\Theta_e$  are replaced  by $\Psi_o$ and
$\Theta_o$. Therefore, we must also have 
$\int_{\Omega} s({\bf x}) \overline{\Psi}_o \cdot \Phi \, d {\bf x} =0$
and $\int_{\Omega} s({\bf x}) \overline{\Theta}_o \cdot \Phi \, d {\bf x} =
0$, or 
\begin{equation}
\label{eq:s_sym4} \int_{\Omega} \left[ s({\bf x}) - s(x,y,-z) \right]
\overline{\Psi}_o \cdot \Phi \, d {\bf x}  =  0
\end{equation}
and
\begin{equation}
\label{eq:s_sym5}  \int_{\Omega} \left[
  s({\bf x}) - s(x,y,-z) \right] \overline{\Theta}_o \cdot \Phi \, d {\bf x} = 0.
\end{equation}
If $s({\bf x})$ does not satisfy any one of Eqs.~(\ref{eq:s_sym2}) -
(\ref{eq:s_sym5}), then the perturbation process fails at the first
step. 
If $s({\bf x})$  satisfies Eqs.~(\ref{eq:s_sym2}) - (\ref{eq:s_sym5}),
then $\alpha_1$, $\beta_1$ and $k_1$ are real, and $\Phi_1$ decays to zero
exponentially as $|z| \to +\infty$.  At the second step ($j=2$), in
order to obtain real $\alpha_2$, $\beta_2$ and $k_2$, $s({\bf x})$ must satisfy
extra conditions involving $\Phi_1$. To carry out the perturbation
process successfully for all steps, $s({\bf x})$ must satisfy an infinite sequence of
conditions. Therefore, if $s({\bf x})$ does not satisfy symmetry condition 
(\ref{eq:symm1}), it is unlikely for the the perturbed structure
to have a BIC. In that case, the BIC of the original unperturbed
structure is turned to a resonant mode with a finite $Q$-factor. 

\section{Numerical examples}

\begin{figure}[htp]
\centering 
\includegraphics[scale=0.5]{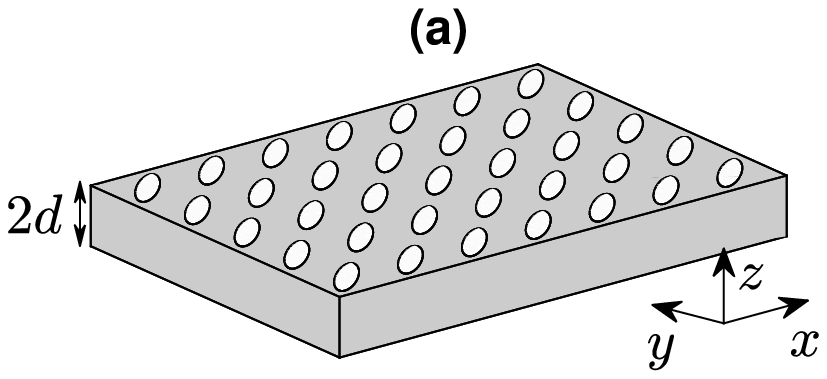}
\includegraphics[scale=0.5]{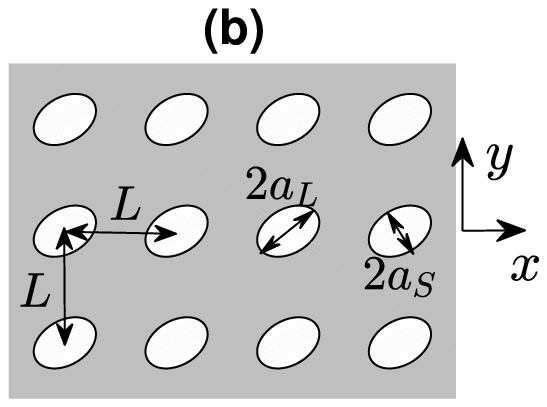}
\caption{A PhC slab with a square lattice of elliptic air holes: (a) 3D view; (b) top view.}
\label{fig:structure}
\end{figure}

To validate  our theory, we consider a photonic crystal (PhC) slab with a square lattice
of elliptic air holes, and demonstrate  the continual
existence of BICs as some structural parameters are varied. As shown in
Fig.~\ref{fig:structure},
the structure is periodic in $x$ and $y$
with period $L$, the thickness  of the slab is $2d=0.5L$,  the
semi-major and semi-minor axes of the elliptic air holes are 
$a_L$ and $a_S$, respectively, and the angle between
the major axis and the $x$ axis is $\theta = \pi/6$. In addition, the dielectric
constant of the PhC slab is $\varepsilon_1$ and the medium surrounding
the PhC slab is air. 

\begin{figure}[htp]
\centering 
\includegraphics[scale=0.78]{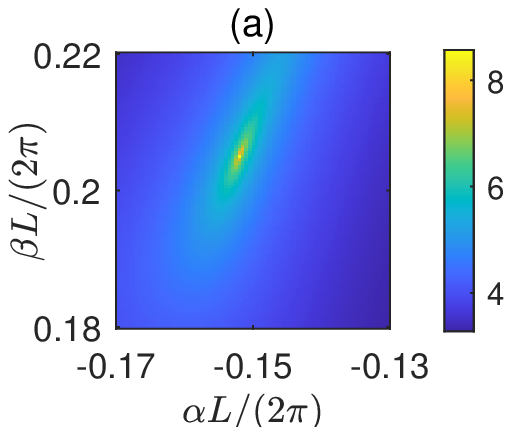}
\includegraphics[scale=0.75]{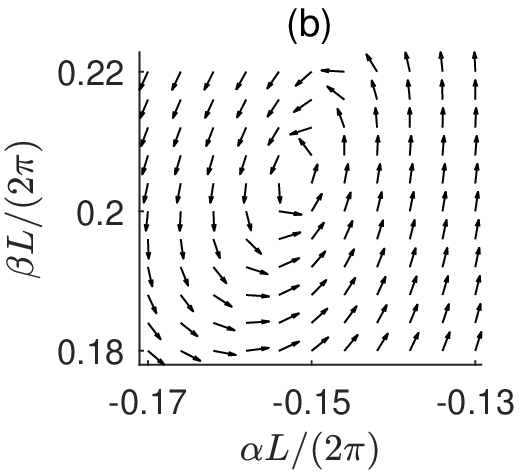}
\includegraphics[scale=0.8]{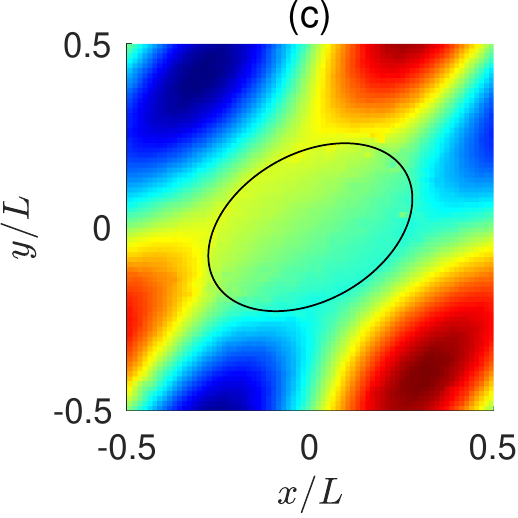}
\includegraphics[scale=0.8]{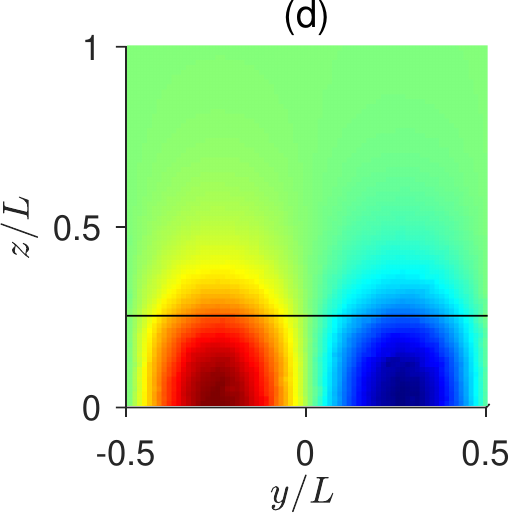}
\caption{A BIC and its nearby resonant modes on a PhC slab with
  elliptic air holes ($\varepsilon_1 = 4$,  $a_L=0.3 L$, $a_S=0.2L$).
  (a) logrithmic value of the $Q$ factor, $\log_{10} Q$, of the
  resonant modes; (b) polarization direction of 
  resonant modes; (c) and (d):
  the imaginary part of $H_z e^{- i (\alpha_* x + \beta_* y)}$ of the BIC on planes at $z=0$ and
  $x=0.5L$, respectively. Also shown in (c) and (d) are the boundary
  of the elliptic air hole and the horizontal slab-air interface.}
\label{fig:BIC}
\end{figure}

For $\varepsilon_1=4$, $a_L=0.3 L$ and $a_S=0.2L$, the PhC slab has a
few symmetry-protected BICs and a few propagating BICs. A particular 
TM-like propagating BIC, of which $E_z$ is odd in $z$ and $H_z$ is
even in $z$, has a Bloch wavevector $(\alpha_*, \beta_*)  = (-0.1515,
0.2057) (2\pi/L)$ and a frequency $\omega_* = 0.7017 (2\pi c/L)$.  In 
Figs.~\ref{fig:BIC}(a) and \ref{fig:BIC}(b),
we show the $Q$ factor
and the polarization direction of the resonant modes near this
BIC, i.e., for $(\alpha, \beta)$ near $(\alpha_*, \beta_*)$.
The polarization direction of a resonant mode is the direction of the
major axis of the polarization ellipse in the far field \cite{bulg17pra4}. The BIC
is the center of the polarization vortex in the $\alpha$-$\beta$ plane. As
$(\alpha, \beta)$ traverses in the counterclockwise  direction along a closed curve 
encircling $(\alpha_*, \beta_*)$, the continuously defined polarization direction
accumulates a total angle of $2\pi$.  Therefore, the topological
charge of this BIC is $q=+1$ \cite{bulg17pra4}.
In Figs.~\ref{fig:BIC}(c) and \ref{fig:BIC}(d), we show the imaginary
part of $H_z e^{- i (\alpha_* x + \beta_* y)}$ on a horizontal plane at $z=0$ and a vertical plane at
$x=0.5L$, respectively. 

\begin{figure}[htp]
\centering 
    \includegraphics[scale=0.51]{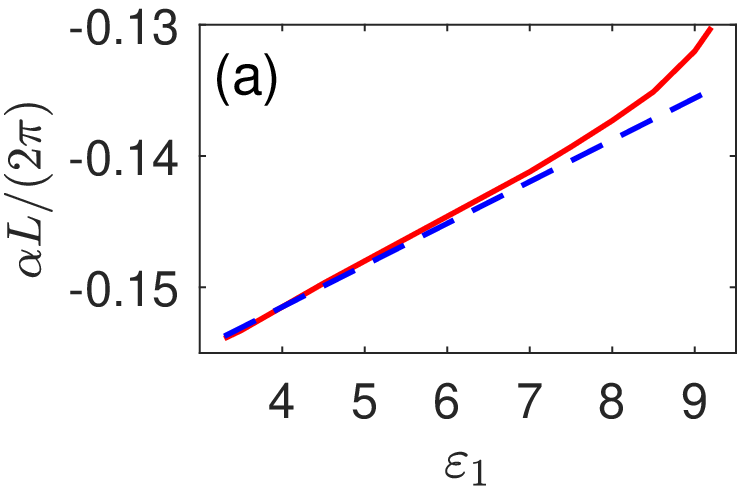}
    \includegraphics[scale=0.51]{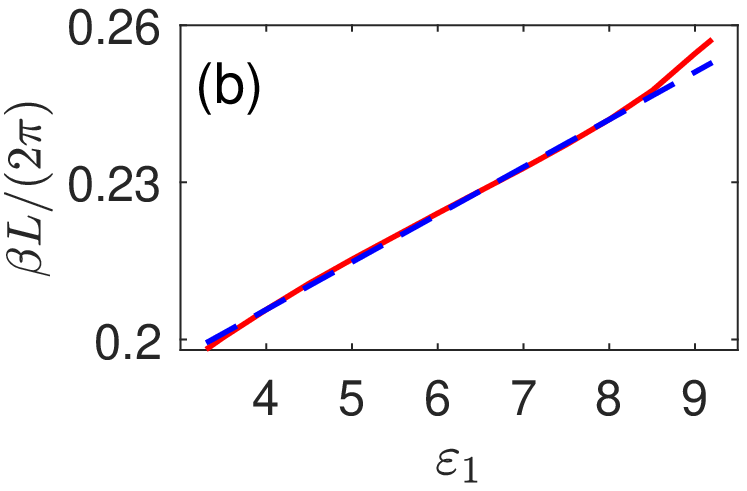}
    \includegraphics[scale=0.51]{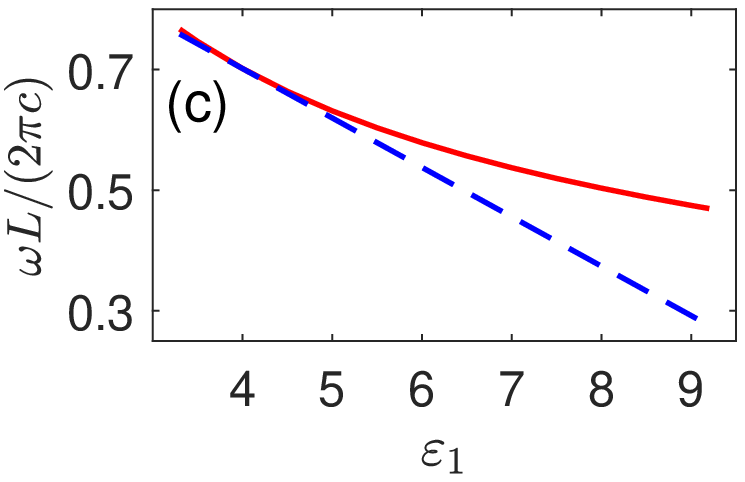}
    \caption{Bloch wavevector $(\alpha, \beta)$ and frequency $\omega$
      of a propagating BIC for different values of
      $\varepsilon_1$. The  red solid lines and the  blue dashed lines are the direct
      numerical results and first order approximations, respectively.}
\label{fig:Example1}
\end{figure}

\begin{figure}[htp]
\centering 
    \includegraphics[scale=0.74]{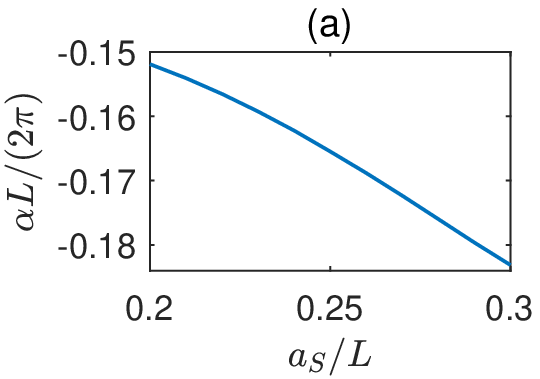}
    \includegraphics[scale=0.74]{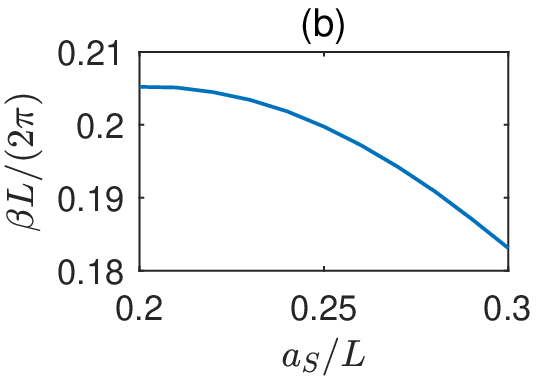}
    \includegraphics[scale=0.74]{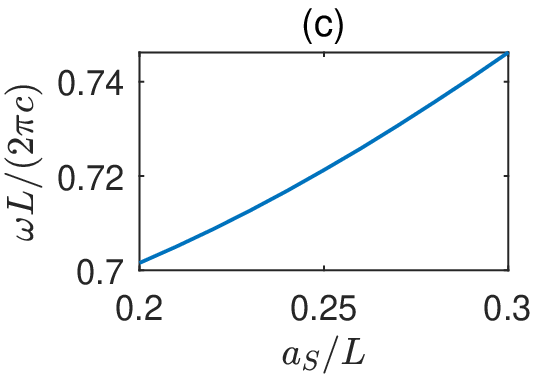}
\caption{Bloch wavevector $(\alpha, \beta)$ and frequency $\omega$ of
  a propagating BIC for different values of $a_S$.}
\label{fig:Example2}
\end{figure}

According to our theory, this propagating BIC is robust against structural perturbations
satisfying the symmetry condition (\ref{eq:symm1}). Our numerical
results confirm that the BIC exists as a continuous family when $\varepsilon_1$ and
$a_S$ are varied. 
In Fig.~\ref{fig:Example1},
we show $\alpha$, $\beta$ and $\omega$ of
this BIC family as functions of $\varepsilon_1$ (solid red lines)
for fixed $a_L=0.3L$ and $a_S = 0.2L$. Useful approximations can be
obtained by keeping only the first order terms in the perturbation
theory. Assuming the unperturbed structure is the PhC slab with
$\varepsilon_1=4$, then the perturbation of the dielectric function is
$\delta s(\mathbf{x})$, where $\delta = \varepsilon_1 - 4$,
$s(\mathbf{x}) = 1$ for $\mathbf{x}$ in 
the slab but outside the air holes and $s(\mathbf{x}) = 0$ otherwise.
The first order perturbation terms can be calculated numerically, and
they are $\alpha_1 = 0.0032 (2\pi/L)$, $\beta_1 = 0.0091 (2\pi/L)$,
and $k_1 = -0.0820 (2\pi/L)$. In Fig.~\ref{fig:Example1}, the
first order approximations, i.e., $\alpha \approx \alpha_* + \alpha_1 (\varepsilon_1 -
4)$,   $\beta \approx \beta_* + \beta_1 (\varepsilon_1 - 4)$, and $k \approx
k_* + k_1 (\varepsilon_1 - 4)$,  are shown as the blue dashed lines. The
results are quite accurate for $\varepsilon_1$ near $4$. 
%
The propagating BIC also  exists continuously with respect to $a_L$
and $a_S$. In Fig.~\ref{fig:Example2},
we show $\alpha$, $\beta$ and $\omega$ of this BIC family as
continuous functions of $a_S$. These results are obtained for 
fixed $\varepsilon_1=4$ and $a_L=0.3L$.

\section{Conclusion}

For biperiodic structures with inversion and reflection symmetries, we
showed that the low-frequency propagating BICs (with only one radiation
channel) are robust against structural
perturbations that preserve the same symmetries. 
The robustness is only conditional, since a BIC can be easily
destroyed by a general perturbation. So far, we have
assumed that both the original and the
perturbed structure are lossless. It is not difficult to see that the
result is still valid if the perturbation profile  is 
symmetric in $z$ and $\mathcal{PT}$-symmetric in $x$ and $y$, i.e. $s(\mathbf{x}) = s(x,y,-z) = 
\overline{s}(-x,-y,z) = \overline{s}(-\mathbf{x})$. 
%
%
A small material loss can also be considered as a perturbation. It is
obvious that a lossy biperiodic structure cannot have a BIC with 
a real frequency and a real Bloch wavevector. It turns out that it
usually cannot even have a bound state with a complex frequency and a
real non-zero Bloch wavevector  \cite{hu20pra}. In other words, if the
original lossless structure has a propagating BIC  and the perturbation profile $s({\bf
  x})$ represents material loss and  satisfies the symmetry
condition (\ref{eq:symm1}), then a propagating BIC is usually
destroyed by the perturbation. 

The theory developed in this paper is applicable to symmetry-protected
BICs, but the reflection symmetry in $z$ is not necessary. Assuming
the biperiodic structure has only the inversion symmetry in 
the $xy$ plane, i.e. $\eps({\bf x}) = \eps(-x,-y,z)$, a
symmetry-protected BIC is a standing wave (with a zero Bloch
wavevector) satisfying 
\begin{eqnarray*}
 E_x({\bf x}) &=& -E_x(-x,-y,z), \quad E_y( {\bf x}) = -E_y(-x,-y,z), \\
 E_z({\bf    x}) &=& E_z(-x,-y,z). 
 \end{eqnarray*}
If we follow the scaling process of Sec.~IV, then the electric field
$E$ of the symmetry-protected BIC is pure
imaginary.  Meanwhile, we can construct two real diffraction solutions
$U^{(1)}(\mathbf{x})$ and $U^{(2)}(\mathbf{x})$ for normal incident
waves such 
that 
\begin{eqnarray*}
 U^{(l)}_{x}({\bf x}) &=& U^{(l)}_{x}(-x,-y,z),  \quad
U^{(l)}_{y}({\bf x}) = U^{(l)}_{y}(-x,-y,z), \\
  U^{(l)}_{z}({\bf
  x}) &=& -U^{(l)}_{z}(-x,-y,z)
\end{eqnarray*}
for $l=1$ and 2.  The perturbation process of Sec.~V is still valid,
provided that we replace $\Phi$, $\Psi_e$ and $\Theta_e$ by $E$, $U^{(1)}$ and $U^{(2)}$, 
respectively.
In each step, we can show that $\alpha_j = \beta_j = 0$, $k_j$ is
real, and 
Eqs.~(\ref{eq:Phi_j3}) and (\ref{eq:Phi_j4}) has a solution $\Phi_j$
that decays to zero exponentially as $|z| \to +\infty$ and satisfies the same symmetry
condition as the BIC.

Finally, we point out that the justification
for conditional robustness presented in this paper is
still somewhat informal. It is desirable to develop a rigorous
proof, including the convergence of the series
(\ref{eq:Pert_Phi}) - (\ref{eq:Pert_beta}),  using a proper
functional analysis framework.

 \section*{Acknowledgements}

 The authors acknowledge support from the Natural Science Foundation
 of Chongqing, China (Grant No. cstc2019jcyj-msxmX0717),  the program for the Chongqing Statistics Postgraduate Supervisor Team (Grant No. yds183002),  and the
 Research Grants Council of Hong Kong Special Administrative Region,
 China (Grant No. CityU 11305518). 
 
\section*{Appendix A}

To derive the first equation of the linear system
(\ref{eq:Linear_sys}), we take the dot product of
$\overline{\Phi}$ with Eq.~(\ref{eq:Phi_j3}), integrate on domain
$\Omega = \left\{(x,y,z) \, : \, |x| < L/2, \, |y|<L/2, \, |z| < +\infty
\right\}$, and obtain
\begin{equation} 
\label{eq:App_Eq1}   \int_{\Omega} \overline{\Phi} \cdot \mathcal{L} \Phi_j d {\bf x} = a_{11} \alpha_j + a_{12} \beta_j + a_{13} k_j - b_{1j}. 
\end{equation}
We need to show that the left hand side above is zero. Since 
$\mathcal{L} \Phi = 0$, we have $\int_{\Omega} \Phi_j \cdot 
\overline{\mathcal{L}} \overline{\Phi} d {\bf x} = 0$, and thus  
\begin{eqnarray}
\label{eq:App_Eq1_Left} 
\nonumber 
&&  \int_{\Omega} \overline{\Phi} \cdot \mathcal{L} \Phi_j d {\bf
             x}  =  \int_{\Omega} \overline{\Phi} \cdot  \left( \nabla  \times  \nabla
   \times \Phi_j \right) d {\bf x} \nonumber \\
  && + i \int_{\Omega}  \overline{\Phi}
   \cdot \left[    \nabla  \times \left( \mathbf{b}  \times \Phi_j
   \right) \right] d {\bf x}  + i \int_{\Omega}  \overline{\Phi} \cdot
   \left[   \mathbf{b}   \times \left(   \nabla\times \Phi_j \right)
   \right] d {\bf x} \nonumber \\
   & & - \int_{\Omega}  \overline{\Phi} \cdot \left[   \mathbf{b}   \times
   \left( \mathbf{b} \times \Phi_j \right) \right] d {\bf x}   -  \int_{\Omega} {\Phi_j} \cdot  \left( \nabla  \times  \nabla \times
  \overline{\Phi} \right) d {\bf x} \nonumber \\
  & & + i \int_{\Omega}  \Phi_j \cdot
  \left[    \nabla  \times \left( \mathbf{b}  \times \overline{\Phi}
  \right) \right] d {\bf x}   + i \int_{\Omega}  \Phi_j \cdot \left[
  \mathbf{b}   \times \left(   \nabla \times \overline{\Phi} \right)
  \right] d {\bf x} \nonumber \\
  & & + \int_{\Omega}  \Phi_j \cdot \left[   \mathbf{b}
  \times \left( \mathbf{b} \times \overline{\Phi} \right) \right] d
  {\bf x}.  
 \end{eqnarray}
Using the vector identities
\begin{eqnarray}
\label{vi1}
&& A \cdot \left( \nabla \times B \right) =  B \cdot \left( \nabla
   \times A \right)  + \nabla \cdot \left( B \times A \right), \\
\label{vi2}
&& A \cdot \left( B \times C \right) = B \cdot \left( C \times A
   \right)  = C \cdot \left( A \times B \right)
\end{eqnarray}
in Eq.~(\ref{eq:App_Eq1_Left}), we have
\begin{eqnarray} 
\label{eq:App_Eq1_1} & &\nonumber \int_{\Omega} \overline{\Phi} \cdot \mathcal{L} \Phi_j d {\bf x}  =
 \int_{\Omega}  \nabla \cdot \left[ \left(\nabla \times \Phi_j \right) \times \overline{\Phi} \right]  d {\bf x} \\ 
 & & + i \int_{\Omega}  \nabla \cdot \left[ \left(\mathbf{b} \times \Phi_j \right) \times \overline{\Phi} \right]  d {\bf x} -  \int_{\Omega}  \nabla \cdot \left[ \left(\nabla \times
     \overline{\Phi} \right) \times \Phi_j \right]  d {\bf x} \nonumber \\
    & & + i
     \int_{\Omega}  \nabla \cdot \left[ \left(\mathbf{b} \times
     \overline{\Phi} \right) \times \Phi_j \right]  d {\bf x}. 
\end{eqnarray}
Because of the Gauss' Law, the above equation becomes
\begin{eqnarray}
& & \nonumber \int_{\Omega} \overline{\Phi} \cdot \mathcal{L} \Phi_j d {\bf x}  = 
\int_{\partial \Omega} \left[ \left(\nabla \times \Phi_j \right) \times \overline{\Phi} \right] \cdot d \mathbf{S} \\ 
& & + i \int_{\partial \Omega} \left[ \left(\mathbf{b} \times \Phi_j \right) \times \overline{\Phi} \right] \cdot d \mathbf{S} - \int_{\partial \Omega} \left[ \left(\nabla \times \overline{\Phi} \right) \times \Phi_j \right] \cdot d \mathbf{S} \nonumber  \\ 
&& + i \int_{\partial \Omega} \left[ \left(\mathbf{b} \times \overline{\Phi} \right) \times \Phi_j \right] \cdot d \mathbf{S} . 
 \end{eqnarray}
Since $\Phi$ and $\Phi_j$ are periodic in the $x$ and $y$ directions, and $\Phi $
decays to zero exponentially as $|z| \to \infty$, the surface
integrals above are zero. Therefore, $\int_{\Omega} \overline{\Phi} \cdot
\mathcal{L} \Phi_j d {\bf x} = 0$.  

Since we assumed that $\Phi_j$ decays to zero exponentially as $|z| \to
\infty$, $\overline{\Psi}_e \cdot {\cal L} \Phi_j$ 
and $\overline{\Theta}_e \cdot {\cal L} \Phi_j$ are also integrable on
$\Omega$. Using the same steps above, it can be shown that their
integrals are zero. This leads to the second and third equations in
system (\ref{eq:Linear_sys}).

\section*{Appendix B}
Unlike the case considered in Appendix A, we only know $\Phi_1$ is
outgoing as $z \to \pm \infty$. This implies that $\Phi_1$ is bounded at infinity,
and we have to consider the integrals on a bounded domain
$\Omega_h$ first. 
To derive Eq.~(\ref{eq:coeff_phi1}), we note that $\mathcal{L} \Psi_e
= 0$. Therefore,  
$$  \int_{\Omega_h} \overline{\Psi}_e \cdot \mathcal{L} \Phi_1 d {\bf
  x} = \int_{\Omega_h}   \left( \overline{\Psi}_e \cdot \mathcal{L}
  \Phi_1 - \Phi_1 \cdot \overline{\mathcal{L}} \overline{\Psi}_e
\right)  d {\bf x}.  $$
Using vector identities (\ref{vi1}) and (\ref{vi2}) and Gauss' Law,  we obtain
\begin{eqnarray*}
&&  \int_{\Omega_h} \overline{\Psi}_e \cdot \mathcal{L} \Phi_1 d {\bf x}  \\
&& = \int_{\partial \Omega_h} \left[ (\nabla \times \Phi_1) \times
    \overline{\Psi}_e - (\nabla \times \overline{\Psi}_e ) \times
    \Phi_1  \right] \cdot d \mathbf{S} \nonumber \\ 
   && + \int_{\partial \Omega_h}  i \left[ ( \mathbf{b} \times \Phi_1 ) \times
   \overline{\Psi}_e +  (\mathbf{b} \times \overline{\Psi}_e) \times
   \Phi_1 \right] \cdot d \mathbf{S}.
\end{eqnarray*}
Since $\Phi_1$ and $\overline{\Psi}_e$ are periodic in the $x$ and
$y$ directions, the integral on the surface parallel
to the $z$-axis is zero.  Thus 
\begin{eqnarray*}
 && \int_{\Omega_h} \overline{\Psi}_e \cdot \mathcal{L} \Phi_1 d {\bf x}  = \int_{D_{xy}}  \mathbf{e}_3 \cdot \left[ (\nabla \times \Phi_1)
  \times \overline{\Psi}_e - (\nabla \times \overline{\Psi}_e ) \times \Phi_1 \right] d {\rm r}  \nonumber \\
  & & + i  \int_{D_{xy}}  \mathbf{e}_3 \cdot \left[( \mathbf{b} \times \Phi_1 ) \times \overline{\Psi}_e + 
  (\mathbf{b} \times \overline{\Psi}_e) \times \Phi_1
  \right]^{z=h}_{z=-h} d{\rm r}, 
\end{eqnarray*}
where  ${\rm r}=(x,y)$, $D_{xy} = \left\{ (x,y) \, : \,  |x| < L/2,
  \,   |y| < L/2\right\}$, and $\mathbf{e}_3 = (0, 0, 1)$ is the unit vector along the $z$ axis.  Based in the asymptotic formulae
(\ref{eq:E_s_e}) and (\ref{eq:Phi_1}), it is not difficult to show that 
$$ \lim\limits_{h \to +\infty} \int_{\Omega_h} \overline{\Psi}_e \cdot \mathcal{L} \Phi_1 d {\bf x} = - 2 i \gamma L^2 \left( \mathbf{c}^+ \cdot \mathbf{d}^+ +  \mathbf{c}^- \cdot \mathbf{d}^- \right).  $$
Noting that $\mathbf{c}^- \cdot \mathbf{d}^- = \mathbf{c}^+ \cdot
\mathbf{d}^+$, we obtain Eq.~(\ref{eq:coeff_phi1}).

\end{document}